\documentstyle[12pt,draft]{article}
\begin{document}
\begin{center}
\Large \bf
Study of non-equilibrium effects and thermal properties of heavy
ion collisions using a covariant approach
\footnote{Supported
by the GSI-Darmstadt and the BMFT under contract 06T\"U736.} \\
\vspace*{1.0cm}
\normalsize \bf\large
Rajeev K. Puri, E. Lehmann, Amand Faessler and S.W. Huang\\
\normalsize\it
\vspace*{0.5cm}
Institut f\"ur Theoretische Physik der Universit\"at T\"ubingen,\\
 Auf der Morgenstelle 14, D72076, T\"ubingen, Germany.\\
\vspace*{1.7cm}
\today
\vspace*{1.5cm}
\bf\normalsize
{\bf Abstract:}\\
\vspace*{0.8cm}
\begin{minipage}{12.5cm}
Non-equilibrium effects are studied using a full
Lorentz-invariant formalism. Our analysis shows that in
reactions considered here, no global or local equilibrium is
reached. The heavier masses are found to be equilibrated more
than the lighter systems. The local temperature is extracted using hot
Thomas Fermi formalism generalized for the case of two
interpenetrating pieces of nuclear matter. The temperature is
found to vary linearly with bombarding energy and impact
parameter whereas it is nearly independent of the mass of the colliding
nuclei. This indicates that the study of temperature with medium
size nuclei is also reliable. The maximum temperatures obtained
in our approach are in a
nice agreement with  earlier calculations of other approaches. A
simple parametrization of maximal temperature as a function of
the bombarding  energy is also given.
\end{minipage}
\end{center}
\newpage
\baselineskip 24 pt
Heavy ion reactions from few MeV/nucl. to few GeV/nucl. provide
an unique opportunity to study the non-equilibrium dynamics of
finite size systems and also the thermal properties of hot and
dense nuclear matter formed during the reaction. Unfortunately,
the hot and dense nuclear matter exits only for very short time.
 In addition, there is no direct way to measure it and
thus various theoretical models are very helpful in the sense
that  one can
simulate the reaction from the start up to the final stage where experimental
measurements are done. The intermediate energy
heavy ion reactions show a non-equilibrium situation in momentum
space. Therefore, for a reasonable understanding of the heavy ion
dynamics, one has to take care of these non-equilibrium effects.

There have been some attempts to extract the temperature
 reached during the heavy ion reactions. Hahn and
St\"ocker \cite{hahn1986} extracted the temperature
 from measured pion yields and got reasonable temperature of the
system at the time of maximum compression. Later on,   there were
some attempts to extract the thermal properties of heavy ion
collisions using dynamical models like Quantum Molecular
Dynamics [QMD] \cite{berenguer1992} and Relativistic
Boltzmann-Uehling-Uhlenbeck
[RBUU] \cite{lang1990},
\cite{lang1991}. These models include the non-equilibrium features. In
the study using RBUU, one assumes that the thermalization is directly
connected with the nondiagonal elements of the stress tensor
and the pressure P can be calculated as the trace of this
tensor. Therefore, from a given form of the nuclear equation of
state (EOS), one deduces the temperature as a function of
energy, density and pressure. In this approximation, one can
extract  two components of the temperature which originate
from the transverse and longitudinal components of the stress
tensor \cite{lang1990}, \cite{lang1991}.


Very recently, Faessler and collaborators [5-10] developed a
novel approach to extract the thermal properties of hot nuclear
matter formed in heavy ion
reactions. In this approach, one starts from a microscopic
picture of  two interpenetrating pieces of nuclear matter and
deduces the thermal quantities from the density of the nuclear matter and
kinetic energy densities obtained during the collisions. The extraction
of the temperature is based on  a Generalized Hot Thomas Fermi
Formalism [GHTFF] \cite{barranco1981}, \cite{rashdan1987}.
 In our approach, the extraction of thermal
properties of hot nuclear matter formed in heavy ion collisions is
 done in two steps :(i)  First one has to simulate the
heavy ion reaction using a reliable dynamical model which
incorporates the non-equilibrium features and can  generate the
 motion of all particles in phase
space during the whole reaction time. (ii) Second using this
information in phase
space, one can calculate the nuclear matter and kinetic energy
densities at each point in coordinate space and time.
 These matter and kinetic energy densities are used to
extract local temperature, entropy etc. \cite{khoa1992a}.

In  all our previous attempts to study the thermal
  properties of heavy ion reactions [5-10],
   we have simulated  heavy ion collisions using
 Quantum Molecular Dynamics (QMD) in a non-covariant form \cite{aichelin1991}.
 While using a non-covariant formalism, one has to assume that
the size of relativistic corrections is small and
hence one can neglect these corrections appearing from a
covariant treatment. To avoid such
assumption,  in
this paper, we couple our recently reported full covariant QMD
[i.e. Relativistic Quantum Molecular Dynamics {RQMD}]
\cite{lehmann1994}, \cite{puri1994c} with the hot
Thomas Fermi Formalism and study the thermal properties of heavy
ion collisions. In the following, we will first give
 a brief introduction of the
formalism used in RQMD and then  present our results
about non-equilibrium effects and temperatures
 reached in heavy ion collisions.

The covariant generalization of QMD model is done in the
framework of the Constraint  Hamiltonian Dynamics. This approach
dubbed as RQMD describes the propagation of all kinds of
baryons in a Lorentz-invariant fashion.
In this approach,  important quantum
features like Fermi motion of nucleons, respect of
the Pauli-principle etc. are also taken into account.
Here  each nucleon of the colliding nuclei are represented by
Gaussian wave-packets. The center of each Gaussian is chosen at
the start of the reaction by a procedure
which is based on the random choice of
 the positions of the nucleons in coordinate and momentum space. This
is done with the help of a standard Monte-Carlo
procedure.  The nucleons are distributed in each of the two
nuclei in a sphere of radius
R = 1.14 A$^{1/3}$ which is consistent with the liquid drop
model. If the centers of two Gaussians are
closer than a distance D$_{min}$ = 1.5 fm, the choice of these
 coordinates is rejected and other coordinates are chosen.
 The momenta of the nucleons are chosen randomly between
zero and the local Fermi momentum.

In RQMD, the Hamiltonian for an N-particle system is expressed in terms of 8N
variables (4N position  $q_{i\mu}$ and 4N momentum
coordinates $p_{i}^{\mu}$).
This means that each particle carries
its own energy and time. Since the physical events are
described as  the world lines in a 6N dimensional phase-space,
extra 2N-1 degrees of freedom have to be eliminated and a global
evolution parameter $\tau$ has to be defined. This can be
achieved with the help of 2N constraints. In our approach,
first N  constraints are chosen as
Poincar$\acute{e}$ invariant on mass shell constraints
which request
that the particles move on energy shell between the collisions
\cite{lehmann1994}, \cite{puri1994c}, \cite{sorge1989}:
\begin{equation}
        \xi_i
= p_i^{\mu}p_{i\mu} - m_i^2 -\tilde V_i =0 \qquad ; \qquad i=1,...,N.
\end{equation}
Here $\tilde V_i$ are the quasi-potentials.
The next set of constraints are chosen as time constraints.
These time constraints fix the relative times
of all particles and are defined as:
\begin{equation}
\chi_{i} = \sum_{j(\neq i)}
 \frac{1}{q^2_{ij}/L_C}\mbox{exp}(q^2_{ij}/L_C)
\qquad i=1,...,N-1,
\end{equation}
\begin{equation}
\chi_{2N} = \hat{P}^{\mu}Q_{\mu}-\tau = 0.
\end{equation}
with $\hat{P}^{\mu}=P^{\mu}/\sqrt{P^2}$, $P^{\mu}=\sum_i p^{\mu}_i$,
$Q^{\mu}=\frac{1}{N}\sum_i q^{\mu}_i$.

 The Hamiltonian is a linear combination of the
Poincar$\acute{e}$ invariant constraints:
\begin{equation}
     H = \sum_{i=1}^{2N-1}\lambda_i \Psi_i,
\end{equation}
with
\begin{equation}
\Psi_i = \left\{
 \begin{array}{c c l} \xi_i & ; & i \leq N \\
                      \chi_{i-N} & ; & N < i \leq 2N-1. \\
\end{array}
\right.
\end{equation}
The unknown $\lambda_i$ in eq.(4) are determine in each time step  using
the condition that all 2N constraints must be fulfilled during
the whole reaction.  After
solving eq.(4), the final equations of motion reads as
\cite{lehmann1994}, \cite{puri1994c}, \cite{sorge1989}
\begin{equation}
\frac{dq_i^{\mu}}{d\tau} =
       2 \lambda_i p_i^\mu - \sum_{j=1}^N \lambda_j
       \frac{\partial \tilde V_j}{\partial p_{i_\mu}},
\end{equation}
\begin{equation}
\frac{dp_i^{\mu}}{d\tau} =
       \sum_{j=1}^N \lambda_j
       \frac{\partial \tilde V_j}{\partial q_{i_\mu}}.
\end{equation}

The propagation of baryons
is combined with  quantum effects like
 stochastic scattering and Pauli-blocking etc..
 In RQMD, the collision part is treated in a
covariant fashion. Therefore, all quantities which determine
the collision are Lorentz invariant. For more details see refs.
\cite{lehmann1994}, \cite{puri1994c}.

Here we couple the covariant simulation of heavy ion collisions
with  the generalized hot Thomas Fermi formalism. In the present
study, we do not generalize the GHTFF to a relativistic version
and thus use the same GHTFF as reported in earlier papers [5-10]. The
relativistic generalization would mean that we assume two
ellipsoids in momentum space rather than two spheres as taken in
GHTFF [5-10]. In GHTFF, the extraction of temperature
is based on  local density approximation. We define in
 each local volume element of nuclear matter in
coordinate space and at each time a "temperature" by the
diffuse edge of the deformed Fermi-distribution consisting of
two colliding Fermi spheres which is  typical for a
non-equilibrium distribution in heavy ion collisions. For
more details, we refer the reader to ref. \cite{khoa1992a}.

We consider two possible quantities to study  the degree of
equilibrium  reached in heavy ion collisions. The first
quantity is the
anisotropy ratio $<R_a>$ which is defined as
\begin{equation}
<R_a> = \frac{\sqrt{<p_x^2>} + \sqrt{<p_y^2>}}{2 \sqrt{<p_z^2>}}.
\end{equation}
This anisotropy ratio is an indicator of the global equilibrium
of the system. The word "global" is due to fact that this quantity
does not depend on the local position and thus
represents the equilibrium of the whole system. The full global
equilibrium would mean that the values of $<R_a>$ are close to one.

 The second possible quantity is the relative momentum of two
colliding Fermi spheres which
indicates the deviation from a  Fermi sphere and by that from
the   local
equilibrium. The concept of local equilibrium is used by the
hydrodynamical models to simulate the heavy ion reactions \cite{clare1986}. The
local equilibrium is represented by the average relative momentum
$<K_R>$ between two colliding Fermi spheres and is defined  as :
\begin{equation}
<K_R> = <| {\bf P}_P({\bf r}, t) -  {\bf P}_T({\bf r}, t)|/ \hbar
> ,
\end{equation}
where
\begin{equation}
{\bf P}_i ({\bf r}, t) = \frac{\sum_{j=1}^{A_{i}} {\bf p}_j(t) \rho_j
({\bf r}, t)}{\rho_i({\bf r}, t)}, ~~~~~i = 1, 2.
\end{equation}
Here $ {\bf p}_j$ and $\rho_j$ are the momentum and density
experienced by the j-th particle and i stands for either target
or projectile.
The relative momentum $<K_R>$ depends strongly on the local
position $\bf r$ and hence it is an indicator of the local equilibrium.
In the following, the study of local equilibrium, density and
temperature is carried out in a central sphere around the centre
of mass of the two colliding nuclei with radius of 2 fm.
 The averaged values are shown. For a self-consistent analysis,
we study the heavy ion collisions by varying following three different
situations i.e.

 (i) A wide range of bombarding energies. This will give us an
unique possibility to study the different phenomena which govern
the heavy ion dynamics at different energies. At low energies the
heavy ion dynamics is governed be the mean field whereas at
higher energies, the frequent baryon-baryon collisions decide the fate of the
collision dynamics of the heavy ion reactions.

(ii) Second, the variation of the impact
parameter at a fixed bombarding energy. This gives us the
possibility to study  the effect of the position of the
colliding nuclei at a fixed bombarding energy. In case of the
central collisions, nuclear matter is highly compressed
whereas in case of peripheral collisions, there is nearly no
compression. Though, experimentally it is not possible
to extract the impact parameter, there have
been lot of attempts  in recent years
to find the approximate impact parameter for a specific collision.

 (iii) The third one can study heavy ion reactions
 as a function of the mass of the colliding nuclei
 at a fixed bombarding energy
and impact parameter. Heavy nuclei can be compressed far more
than the light one. In order to have a fair study using different
masses, we will often choose the impact parameter which is a
certain fraction of the radii of the nuclei under
consideration. Therefore,  we define a scaled impact parameter
in units of $b_{max}$
which is equal to the radius of the target plus radius of the projectile.
In these units the same impact parameter
provides the possibility to study reactions of
different nuclei with a corresponding geometry.

Fig.1 shows the time evolution of the anisotropic ratio $<R_a>$
for the reaction of $^{40}$Ca -$^{40}$ Ca at an impact parameter of
2 fm  using the soft and the hard EOS's. We choose two typical
bombarding energies i.e. 50 MeV/nucl.  and 1 GeV/nucl..
 One can also see two different phenomena which govern
the process of thermalization at 50 MeV/nucl. and at 1 GeV/nucl..
 Due to lack of free phase space at 50 MeV/nucl.
 about 80-85 $\%$
of the attempted collisions are blocked \cite{puri1994c}. In other words, the
mean field dominates the collision dynamics at 50 MeV/nucl. and
hence one sees a smooth rise of the $<R_a>$ value from 0.5 to
about 0.8 at 60 fm/c.
 When one sees the lower part of the figure where
simulations at 1 GeV/nucl. are shown, one recognizes that as
soon as two nuclei touch each other, the $<R_a>$ ratio increases
suddenly. Whereas when we take a Vlasov-mode ( i.e. we suppressed
all collisions by definition), we see that there is no sudden change in
the $<R_a>$ value and the final value is about the same as
 at the start of the reaction. The evolution of the anisotropic
ratio at 50 MeV/nucl. and 1 GeV/nucl. in the Vlasov-mode is
quite similar : both show a smooth dependence.
Therefore, this figure demonstrates the two different processes
which are responsible for the thermalization at low and high
energies. At low energies, the mean field is responsible for the
equilibration  whereas at higher energies the
collisions are essential for the rise in the anisotropy ratio.

The dependence of the anisotropy ratio on the mass  of the colliding nuclei
is shown in the upper part of fig.2. The  impact
 parameter is b = 0.25 $b_{max}$. It is interesting to note
that the heavier nuclei are able to equilibrate more than the
lighter nuclei. This can be understood on the ground that the
number of collisions per nucleon for the $^{40}$Ca -$^{40}$Ca
reaction is larger  than
for the $^{12}$C -$^{12}$C reaction. These collisions are responsible
 for thermalization.
If one extrapolate this result to heavy systems, one can
assume that for very heavy masses, one may be able to get nearly
global equilibrium. Further, we also see that the thermalization
process starts a  bit later for heavier nuclei than for
the lighter ones. From the above analysis, it is clear that for the reaction
considered here, no global equilibrium is reached.
 The anisotropy ratio $<R_a>$ reached in all
reactions considered here is less than one. It means that the
average kinetic energy in the transverse direction is lower than
that in the longitudinal direction.

We now study the  local equilibrium. In
the following we will concentrate on the bombarding energies up
to 500 MeV/nucl. only.

The time evolution of the relative momentum $<K_R>$ is shown in
lower part of fig. 2 using the hard EOS. Here five systems
$^{40}$Ca -$^{40}$Ca, $^{28}$Si -$^{28}$Si, $^{20}$Ne -$^{20}$
Ne, $^{16}$O -$^{16}$O, $^{12}$C-$^{12}$C are considered at
an impact parameter b = 0.25 $b_{max}$. It is interesting
to see that the relative momentum is very large at the start of
the reaction and finally at the end of the reaction, the
value of $<K_R>$ is nearly zero. This means that at the end of
the reaction, the local equilibrium is nearly reached.
 This can also be  due to fact that the matter
density at this last phase of the reaction is  very small
(see e.g. fig.6).  One also sees that at the start of the
reaction, heavier masses show smaller values of $<K_R>$. This is due
to fact that in  a covariant approach, we have Lorentz-contracted
initial distributions in coordinate space which increases the
distance of  surfaces of heavier nuclei more at the start of the reaction as
compared to lighter nuclei.  The time evolution of
relative momentum at different energies and impact parameters is
shown in fig.3. It is interesting to
note that at the end of the reaction, the relative average momentum for
higher energies is less than for lower energies. ( At lower energies
most of the two-body collisions are Pauli-blocked and
equilibration is slower).
The variation of the relative momentum with  impact parameter
is more interesting. The starting value of the $<K_R>$ is nearly
the same for all impact parameters,  but  the time
evolution of $<K_R>$ is quite different for central and peripheral
collisions. From the above study, it is
clear that for all energies and reactions considered here, the
local equilibrium is reached only at the last phase of the
reactions. This , however, questions the validity of the hydrodynamical
models for studying  the heavy ion reactions at intermediate
energies. All these results are in nice agreement with
earlier calculations with non-covariant QMD \cite{khoa1992a}.

The evolution of the local temperature and matter density is shown in
fig. 4 for the reaction of  $^{40}$Ca -$^{40}$Ca at an impact
parameter of 2 fm. One should note that the local temperature
depends on the matter and kinetic energy densities. We here
display the results for temperature at the
bombarding energies E$_{lab}$ = 125 MeV/nucl., 250
MeV/nucl., 375 MeV nucl., and 500 MeV/nucl.. One can see the
following interesting results:

 \fbox{i} Compared to low energies, the temperature and density reaches
the maximal value earlier for higher bombarding energies.
 This is simply due to the fact that the
velocities of particles at low energies is far less than that at
higher energies. In order to have a fair measure, one can
rescale the reaction time with the boost velocity $\beta_{cm}$.
The product [$\beta_{cm}$ ~t] is a kind of a reaction
 distance of the two colliding nuclei.
  We notice that the value of  this product $\beta_{cm}$~t
when temperature and density reaches its maximum value is nearly
the same for all energies.

\fbox{ii} The size of  the hot and dense zone depends strongly
on the energy considered. One sees that at low energies, the hot
and dense zone exits for a longer time than at higher energies.

\fbox{iii} The maximal value of the temperature varies linearly
 with the bombarding energy (see, however, fig.8).
Whereas the increase in matter density with bombarding
energy is very small. This shows that the major factor which
causes the temperature is the kinetic energy i.e. the bombarding energy.

The effect of variation in impact parameter on temperature and
density is studied in fig. 5. Here we simulate  the reaction $^{40}$Ca
-$^{40}$Ca  at the bombarding energy  500 MeV/nucl. using the
hard EOS. It
is interesting to note that at  a fixed bombarding energy, the
temperature and density shows a linear correlation which indicates
that the knowledge of either of them can give us a rough
knowledge of the behaviour of other quantity. This linear behaviour is
understandable. By fixing the bombarding energy, the kinetic
energy is fixed
and thus the variation in matter density shows  a linear effect on
the temperature.
The central collisions show a very high value of the temperature
and density whereas peripheral collisions yield smaller values.
 In the case of central collisions,
we have a large fraction of nucleons which are participants
whereas in the case of peripheral collisions, almost all nucleons
are spectators. One also sees that the variation in
the impact parameter alters not only the maximal value of
the temperature and the density but, it also affects the size of the hot
and dense zone.
 From figs. 4 and 5, it is clear that for studying the temperature
and density, the bombarding energy and the impact parameter play a
central role. Therefore, in fig. 6, we fix the bombarding energy
and  the impact parameter, but vary the mass of the colliding
nuclei. It is impressive to note that when one fixes the impact
parameter and the bombarding energy, the maximal value of the
temperature reached is about the same for all masses i.e. the maximal
value of the temperature is nearly independent of the mass of
colliding nuclei whereas the matter density depends strongly on
the mass of the colliding pair. One can also see that the
temperature and  density reaches the maximal value for later
times in heavier compared to the lighter systems. This different response
of the temperature and the density to the total mass
at a fixed bombarding energy and
impact parameter indicates clearly that for determining the
temperature, the major factor is the bombarding energy rather than
the matter density. This result also shows that for studying
the temperature and thermal properties, one can rely on medium
size nuclei. This result is understandable since
the bombarding energy fixes the possible excitation energy of the system.

The results presented in figs.4-6 are calculated with the hard
EOS only. The influence of different
equation of states on the temperature is studied in fig. 7.
Here simulations of the reaction $^{40}$Ca-$^{40}$Ca are shown at impact
parameters b = 0 and 5 fm, respectively. The soft EOS creates far
more temperatures than the hard EOS. We also note that when one
goes to larger impact parameters, the difference in temperature
using the hard and the soft EOS decreases. It is also worth to mention
 that all calculations presented in this paper are using the
cross-sections parametrized by Cugnon \cite{cugnon1981}. If one does the
calculations using in-medium cross-sections, one finds that the
use of in-medium cross-sections results in significant
enhancement in temperature {\cite{khoa1992b}. Similar enhancement is found when
one uses the cross-section derived from one-boson exchange
model {\cite{puri1994d}.


In fig.8, we plot the maximal value of the average temperature reached
during the reaction using a variety of colliding masses as a function
of the bombarding energy. Here to constraint the colliding
geometry, we fix the impact parameter ( b = 0.25 $b_{max}$)
to semi-central collisions. The maximum value of
the temperature which is independent of the mass of the colliding nuclei
can be parametrized in the following simple form:
\begin{equation}
T^{max} = 0.09466 \cdot E_{lab} + 4.934 ~~~~~\mbox{for} ~~ b =
0.25 b_{max}.
\end{equation}
Our classical nature of the RQMD does not allow us to analyze
the temperature for bombarding energies less than 40 MeV/nucl..
This parametrization of the maximum temperature can be used to
extract the temperature for the bombarding energies between 40
MeV/nucl. and 500 MeV/nucl.. The linear dependence of the
temperature on the bombarding energy is not surprising since the
bombarding energy determines the possible maximum excitation energy.
 Further it is impressive to note the independence of
temperatures on the masses.

These  values of the maximum temperatures are in a nice agreement with the
values obtained in ref. \cite{hahn1986}. In these
calculations, a  thermodynamically consistent theory
was applied to extract the temperature from measured pion
yields.

In this paper we have  coupled our recently developed
microscopic approach for extracting the temperature
 with the full covariant RQMD. At low energies, the mean
field keeps the nuclei together and hence they have time to
develop to  nearly full
equilibrium. At higher energies, the frequent collisions are
responsible for equilibrium and hence for thermalization. At
higher energies no full equilibrium is reached. Further, heavier
colliding nuclei show a better  tendency for equilibrium. We have also
studied the local equilibrium. We find that even at the last
stage of the reaction, no full local equilibrium is reached. The
 local temperature depends strongly on the
bombarding energy and the impact parameter whereas the maximal value
of the temperature at a fixed bombarding energy and collision
geometry (b/b$_{max}$) is nearly independent of the masses of the
colliding nuclei. This behaviour is similar to the  resonance
production or pion production which are also found to be nearly
independent of the masses of the colliding nuclei.  The maximum
temperature which is independent of the masses of the colliding
nuclei agrees with the one extracted from measured pion yields.
Further a simple formula is also given to calculate the
maximum value of the temperature.

\newpage
{\large\bf Figure Captions:}
\vspace*{0.5cm}

{\bf Fig.1} Time evolution of the anisotropy ratio $<R_a>$ as a
function of the reaction time. The reaction under consideration
is  $^{40}$Ca -$^{40}$Ca  at an impact parameter b = 2 fm. The
upper and lower parts of the figure represent results at
 bombarding energies 50 MeV/nucl.
and 1 GeV/nucl.. The dotted line with open circles
shows the results  using  the Vlasov-mode.

\vspace *{0.5cm}

{\bf Fig.2} Time evolution of the anisotropy ratio $<R_a>$ as a
function of the reaction time (upper part). Here we simulate five systems
 $^{40}$Ca -$^{40}$Ca, $^{28}$Si -$^{28}$Si, $^{20}$Ne -$^{20}$Ne,
 $^{16}$O -$^{16}$O and  $^{12}$C-$^{12}$C  at the impact parameter b =
 0.25 $b_{max}$.  The lower part of the figure shows the
averaged relative momentum $<K_R>$ as
a function of the reaction time. The bombarding energy is 500 MeV/nucl..

\vspace*{0.5cm}

{\bf Fig.3} Time evolution of the relative momentum $<K_R>$ as a
function of the reaction time. The reaction under consideration
is  $^{40}$Ca -$^{40}$Ca using the hard EOS. The upper part
of the figure is calculated for the  impact parameter 2 fm and the bombarding
energies  50 MeV/nucl., 125 MeV/nucl., 250 MeV/nucl., 375
MeV/nucl. and 500 MeV/nucl.. The lower part of the figure is for
the bombarding energy 500 MeV/nucl. with impact parameters b =
0 fm, 2.5 fm and 5 fm.

\vspace*{0.5cm}

{\bf Fig.4} Time evolution of the average temperature and matter
density as a function of the reaction time. The upper part shows
the evolution of the temperature at incident energies 125
MeV/nucl., 250 MeV/nucl., 375 MeV/nucl. and 500 MeV/nucl.. The
lower part displays the matter densities at bombarding energies
125 MeV/nucl. and 500 MeV/nucl..

\vspace*{0.5cm}

{\bf Fig.5} The same as in fig. 4 but at the bombarding energy of 500
MeV/nucl. and at impact parameters b = 0 fm, 2.5 fm, 5.0 fm and
7.5 fm.

\vspace*{0.5cm}

{\bf Fig.6} The same as in fig. 5 but at a bombarding energy of 500
MeV/nucl. and at an impact parameter b = 0.25 $b_{max}$.
Here five different systems  $^{40}$Ca -$^{40}$Ca,
 $^{28}$Si -$^{28}$Si, $^{20}$Ne -$^{20}$Ne,
 $^{16}$O -$^{16}$O and  $^{12}$C-$^{12}$C
are considered.

\vspace*{0.5cm}

{\bf Fig.7} The same as in fig.6, but at incident energy of 500
MeV/nucl. The upper and lower parts show the temperature using
the hard and the soft EOS's at impact parameters b = 0 fm and 5 fm,
respectively.

\vspace*{0.5cm}

{\bf Fig.8} The maximal value of the temperature as a function
of the bombarding energy. Here the hard EOS is used. Note that
various colliding systems are considered at an impact parameter
b = 0.25 $b_{max}$. The solid line is the parametrized curve
given by eq.(11)
\end{document}